\documentclass[twocolumn]{jpsj2} 

\title{
The effect of randomness on the quantum spin system Tl$_{1-x}$K$_x$CuCl$_3$ with $\mbox{\boldmath $x$}$ = 0.44 studied by the Zero-field Muon-Spin-Relaxation (ZF-$\mu$SR) method
}
\author{Takao Suzuki$^{1}$\thanks{E-mail address: suzuki\_takao@riken.jp}, Fumiko Yamada$^{2}$, Isao Watanabe$^{1}$, Takayuki Goto$^{3}$,  Akira Oosawa$^{3}$, and Hidekazu Tanaka$^{2}$}
\inst{$^{1}$  Advanced Meson Science Laboratory, RIKEN (The Institute of Physical and Chemical Research), 2-1 Hirosawa, Wako, Saitama 351-0198 \\
$^{2}$ Department of Physics, Tokyo Institute of Technology, Oh-okayama, Meguro-ku, Tokyo 152-8551 \\
$^{3}$ Department of Physics, Sophia University, 7-1 Kioi-cho, Chiyoda-ku, Tokyo 102-8554 
}

\abst{
Zero-field muon-spin-relaxation (ZF-\mbox{\boldmath $\mu$}SR) measurements were carried out down to 80 mK on the randomness bond system Tl$_{1-x}$K$_x$CuCl$_3$ with $x$ = 0.44.
Time spectra are well fitted by the stretched exponential function $\exp (-\lambda t)^{\beta}$.
The muon spin relaxation rate $\lambda$ increases rapidly with decreasing temperature, and $\beta$ tends to 0.5 at 80 mK.
The divergent increase of $\lambda$ suggests the critical slowing down of the frequency of the Cu-3{\it d} spin fluctuations toward a spin frozen state below 80 mK, and the root-exponential-like behavior of the time spectrum indicates that the origin of the relaxation is possibly the spatially-fixed fluctuating dilute moments.

}
\kword{
Tl$_{1-x}$K$_x$CuCl$_3$, spin gap, magnon, Bose-Einstein condensation, randomness, Bose glass phase, $\mu$SR 
}

\begin{document}
\maketitle

\section{Introduction}

\indent
The isostructural materials TlCuCl$_3$ and KCuCl$_3$ are three-dimensionally coupled Cu-3{\it d} S = 1/2 spin dimer systems, and their magnetic ground states are spin-singlets with excitation gaps of $\Delta =$ 7.5 K and 31 K, which originate from strong intradimer antiferromagnetic interaction {\it J}~\cite{Willett,Cavadini,Cavadini2,Takatsu,Shiramura}.
The spin dimers couple with one another through interdimer exchange interactions.
The excited spin triplets (magnons) in the singlet sea can hop to neigbouring dimers due to the transverse component of the interdimer interaction, and the system can be represented as an ensemble of bosonic particles~\cite{Chubukov,Sachdev,Rice,MatsumotoPRL}.
Under strong magnetic fields higher than the gap field {\it H}$_{\rm g} = \Delta /{\rm g}\mu_{\rm B}$, a magnetically ordered state appears, and this field-induced magnetic ordering can be described as Bose-Einstein condensation (BEC) of the excited triplets, because the superfluid and Mott insulating phases of the system of bosonic-particles theoretically correspond to the field-induced ordered phase and the gapped phase in the present spin system~\cite{O3,Oosawa_Tl,Oosawa_K,Higemoto,Nikuni,Tanaka,RueggA,Ruegg,Ruegg2,Matsumoto,Misguich,Goto}.
\\
\indent
In low-dimensional spin gap systems, theoretical studies predict that the ground state can be affected by an exchange bond-randomness~\cite{Dasgupta,DSF}, and the exchange bond-randomness effect on the quantum spin system has been studied experimentally in many materials.
In the random bond one-dimensional Heisenberg chain system (CH$_3$)$_2$CHNH$_3$Cu(Cl$_{x}$Br$_{1-x}$)$_3$, spin gap phases are observed for $x <$0.44 and $x >$0.87, whereas a magnetically ordered phase at zero-field is observed in the concentration region of 0.44 $< x <$ 0.87 by the magnetic susceptibility and specific heat measurements~\cite{Manaka1,Manaka2}.
The long range ordered state was confirmed for $x =$ 0.85 by a muon-spin-rotation measurement~\cite{Got}.
\\
\indent
In the mixed system Tl$_{1-x}$K$_x$CuCl$_3$, which is the subject of this study, the randomness of the local potential is introduced through the difference of the value of the dominant intradimer interaction {\it J} between TlCuCl$_3$ and KCuCl$_3$, because {\it J} corresponds to the local potential of magnons.
Magnetization measurements suggest that the ground state is negative with no gap in the mixed system in zero field (ZF), although both the parent materials are non-magnetic having finite excitation gaps~\cite{Oosawa_TlK}.
In other words, the randomness induces a finite magnetic density of states within the excitation gap.
\\
\indent
The appearance of a new phase, Bose-glass phase, in lower magnetic fields at {\it T} = 0 is predicted by theories.~\cite{Fisher,Totsuka,Nohadani}
According to theoretical predictions, the Bose-glass phase is produced between gapped and ordered phases corresponding to Mott insulating and superfluid phases.
By the correspondence between the spin system and the bosonic-particle system, the uniform magnetization {\it M} and the magnetic susceptibility $\chi = \frac{\partial M}{\partial H}$ correspond to the number of the bosons {\it N} and the compressibility $\kappa$.
In the BEC phase, i.e., the superfluid state of bosons, the system is characterized by a finite staggered magnetization perpendicular to the applied magnetic field, whereas the Bose-glass phase is distinguished by the disappearance of the staggered magnetization keeping the magnetic susceptibility finite.
In other word, bosons are localized due to randomness, but there is no gap.
\\
\indent
Shindo {\it et al}. carried out specific heat measurements in magnetic fields, and observed field-induced phase transitions which are described by the Bose-Einstein condensation of triplets of Cu-3{\it d} spins in the mixed system Tl$_{1-x}$K$_x$CuCl$_3$~\cite{Shindo,Tanaka2}.
They discussed the obtained phase diagram in connection with the appearance of the Bose-glass phase.
Bose-glass phenomena have also been studied intensively in other disordered quantum systems, vortex lattices~\cite{Nelson,Ammor} and trapped atoms~\cite{Wang}.
\\
\indent
Recently, we reported the increase of the muon-spin-relaxation rate $\lambda$ at low temperature as well as the NMR-$T_{\rm 1}$ in Tl$_{1-x}$K$_x$CuCl$_3$ with $x =$ 0.20 single crystals, which is possibly a precursor to the Bose-glass phase at {\it T} = 0~\cite{Suzuki,Fujiwara}.
These results are consistent with the theoretical prediction that the Bose-glass phase is expected to appear for $x >$ 0.
However, it is not yet known whether or not another ground state apart from the Bose-glass phase and the gapped state appears when the randomness is enhanced with increasing the concentration of $x$ in the system.
In order to investigate microscopic magnetic properties and to obtain information about the ground state in highly random systems, we carried out the zero-field muon-spin-relaxation (ZF-$\mu$SR) measurements in Tl$_{1-x}$K$_x$CuCl$_3$ with $x =$ 0.44 single crystals.
\\
\indent
In ionic crystals, the positive muon $\mu^{+}$ usually stops near minus ions.
For example, the muon-site in the case of high-{\it T}$_{\rm c}$ cuprates is determined to be near the O$^{-2}$ ions~\cite{musite}.
In Tl$_{1-x}$K$_x$CuCl$_3$, muon spins are expected to be implanted near the Cl$^{-}$ ions.
We expect that muon spins probe information about the Cu-3{\it d} spins through a hyperfine interaction, because it is reported that muon spins successfully detect Cu-3{\it d} spin fluctuations in the other quantum spin systems (CH$_3$)$_2$CHNH$_3$CuCl$_3$ and (CH$_3$)$_2$CHNH$_3$Cu(Cl$_{x}$Br$_{1-x}$)$_3$ which are similar compounds to Tl$_{1-x}$K$_x$CuCl$_3$~\cite{Got,ManakaNew}.

\section{Experimental}

\indent
Single crystals used in this study were grown from a melt by the Bridgman method.
The details of crystal growth are given elsewhere~\cite{Oosawa_TlK}.
The magnetization was measured using a superconducting quantum interference magnetometer (Quantum Design MPMS XL) in the department of physics, Tokyo Institute of Technology.
Zero-field muon-spin-relaxation (ZF - $\mu$SR) measurements were carried out at the RIKEN-RAL Muon Facility in the U.K. using a spin-polarized pulsed positive surface-muon beam with an incident muon momentum of 27 MeV/c.
Forward and backward counters were located on the upstream and downstream sides of the beam direction, which was parallel to the initial muon-spin direction.
The asymmetry parameter was defined as follows:
\[
A(t)=\frac{F(t)-\alpha B(t)}{F(t)+\alpha B(t)}
\]
$F(t)$ and $B(t)$ were total muon events counted by the forward and backward counters at a time $t$, respectively.
The $\alpha$ is a calibration factor reflecting relative counting efficiencies between the forward and backward counters.
The initial asymmetry is defined as $A(0)$.
In this study, the calibration factor $\alpha$ and the background subtraction were taken into account for the data analysis.
The muon-spin-relaxation ($\mu$SR) time spectra were measured down to $T = $ 80 mK using a dilution refrigerator (Leiden cryogenics b. v.).
The incident muon-spin direction was parallel to the b-axis of single crystals.
Cleaved single crystals were attached densely by an Apiezon N grease on a five nines purity silver plate.
The total size of crystals is 25 $\times$ 3 $\times$ 25 mm$^3$.

\section{Results and Discussion}

\indent
Figure 1 shows the temperature dependence of the magnetic susceptibility $\chi = M/H$ in Tl$_{1-x}$K$_x$CuCl$_3$ with $x$ = 0.44 at the magnetic field of 0.1 T.
The reciprocal of the susceptibility $\chi^{-1}$ is shown in the inset of Fig. 1.
Above 60 K, $\chi^{-1}$ has a linear temperature dependence, which means that the spins are weakly coupled in the higher temperature region.
Below 60 K, however, the $\chi$-{\it T} curve deviates from the Curie-Weiss law, and the susceptibility $\chi$ begins to decrease rapidly with decreasing temperature toward a finite value after a broad peak at 30 K.
\begin{figure}
\scalebox{0.5}{\includegraphics*[10mm,83mm][200mm,241mm]{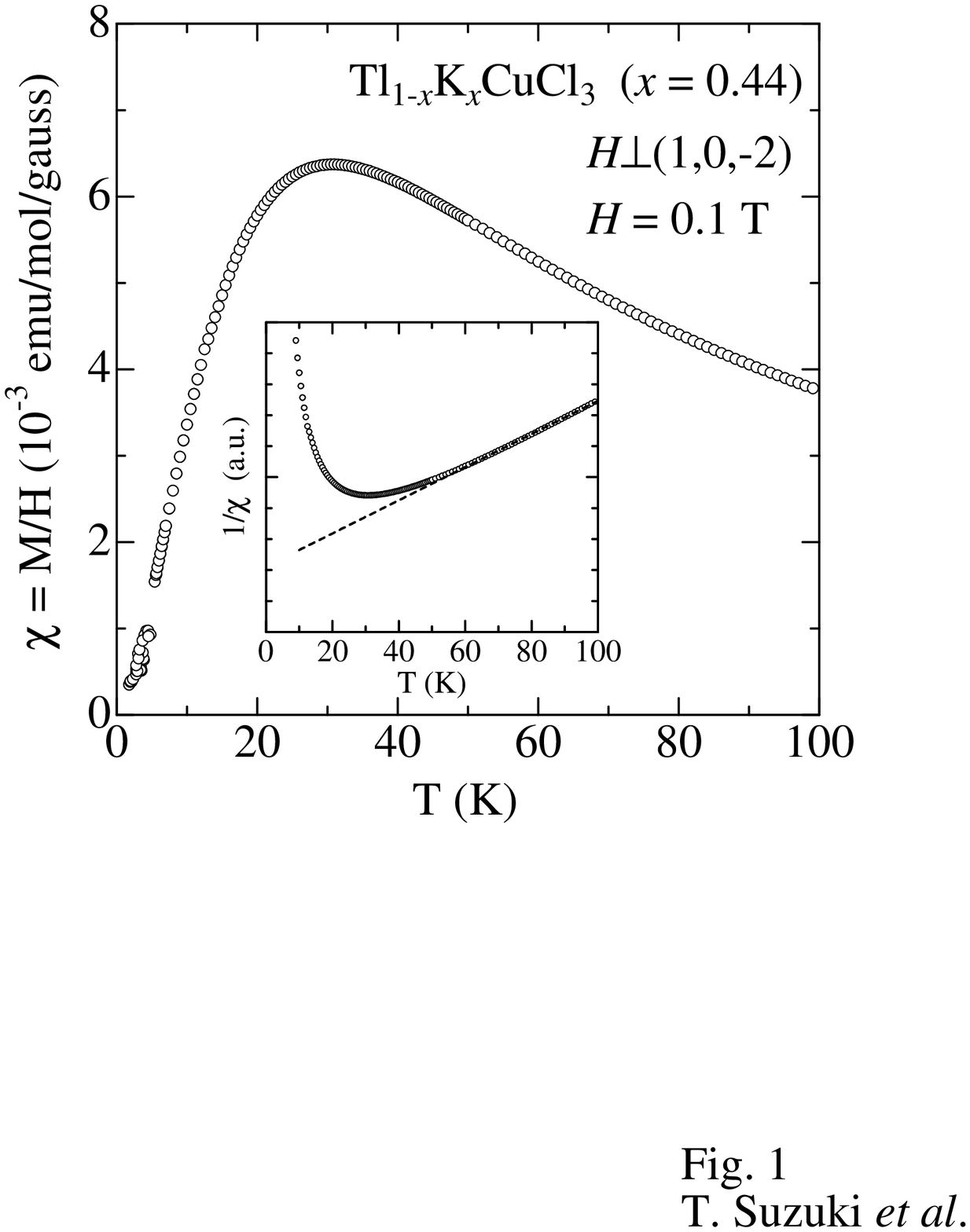}}
\caption{
Temperature dependence of the susceptibility $\chi = M/H$ in the magnetic field of {\it H} = 0.1 T.
The inset shows the temperature dependence of the reciprocal of the susceptibility $\chi^{-1}$.
The dashed line is the fitted result using a linear function in the temperature region between 60 K and 100 K.
}
\end{figure}
The rapid decrease of the susceptibility indicates the development of the antiferromagnetic spin correlations at lower temperatures.
Such temperature dependence of the susceptibility is the characteristic behavior in this system~\cite{Oosawa_TlK,Shindo,Tanaka2}.
\begin{figure}[bbb]
\scalebox{0.5}{\includegraphics*[20mm,94mm][200mm,230mm]{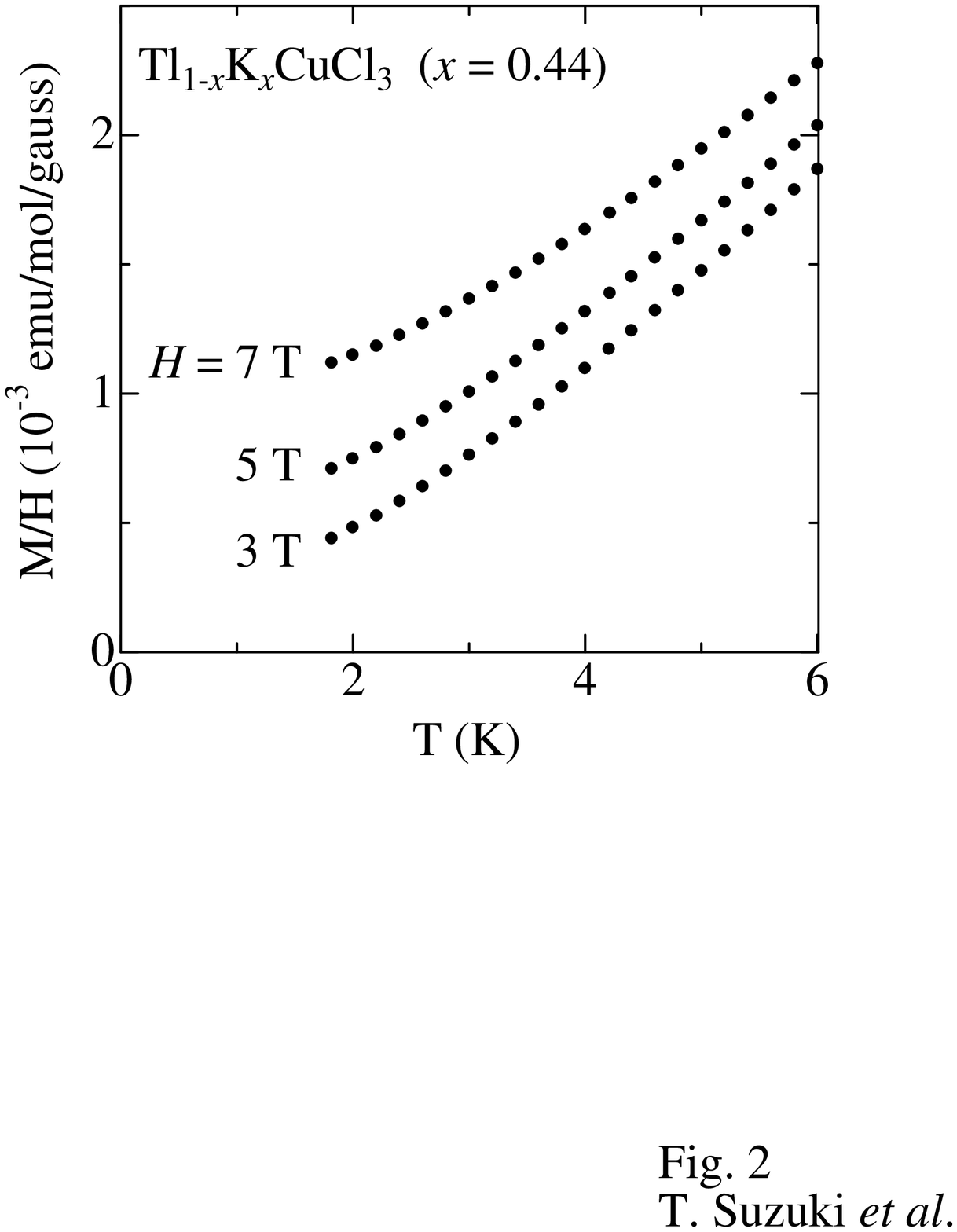}}
\caption{
Temperature dependence of the susceptibility in the magnetic fields {\it H} = 3, 5, 7 T.
}
\end{figure}
\\
\indent
Figure 2 shows the temperature dependence of the magnetic susceptibility at high magnetic fields of 3, 5, and 7 T.
No cusp like minimum is observed down to 1.8 K and up to 7 T, which suggests a shift of the field-induced magnetic phase transition point to lower temperatures and/or to higher magnetic fields region.
This result is consistent with the tendency reported in Tl$_{1-x}$K$_x$CuCl$_3$ with $x < $0.27 (ref.27).
A possible origin of the shift of the transition point is the enhancement of the bond-randomness introduced in the system, because the concentration of $x$ is supposed to correspond to the degree of the randomness.
Another possibility is the development of a uniform spin excitation gap of the system because the excitation gap in KCuCl$_3$ is quite large compared with that in TlCuCl$_3$.
\\
\indent
Figure 3 shows the field dependence of the magnetization ({\it M}-{\it H} curve) at 1.8 K.
A finite value of the differential ${\rm d}M/{\rm d}H$ at the zero-field limit indicates that the ground state is paramagnetic.
As seen in Fig.3, absolute values of the magnetization {\it M} and of the differential ${\rm d}M/{\rm d}H$ are larger compared to those reported in the samples with $x <$ 0.27.
The increase of the magnetization with increasing $x$ means that the density of states for the spin excitations in the low-energy region is enhanced.
\begin{figure}[htb]
\scalebox{0.5}{\includegraphics*[20mm,68mm][200mm,215mm]{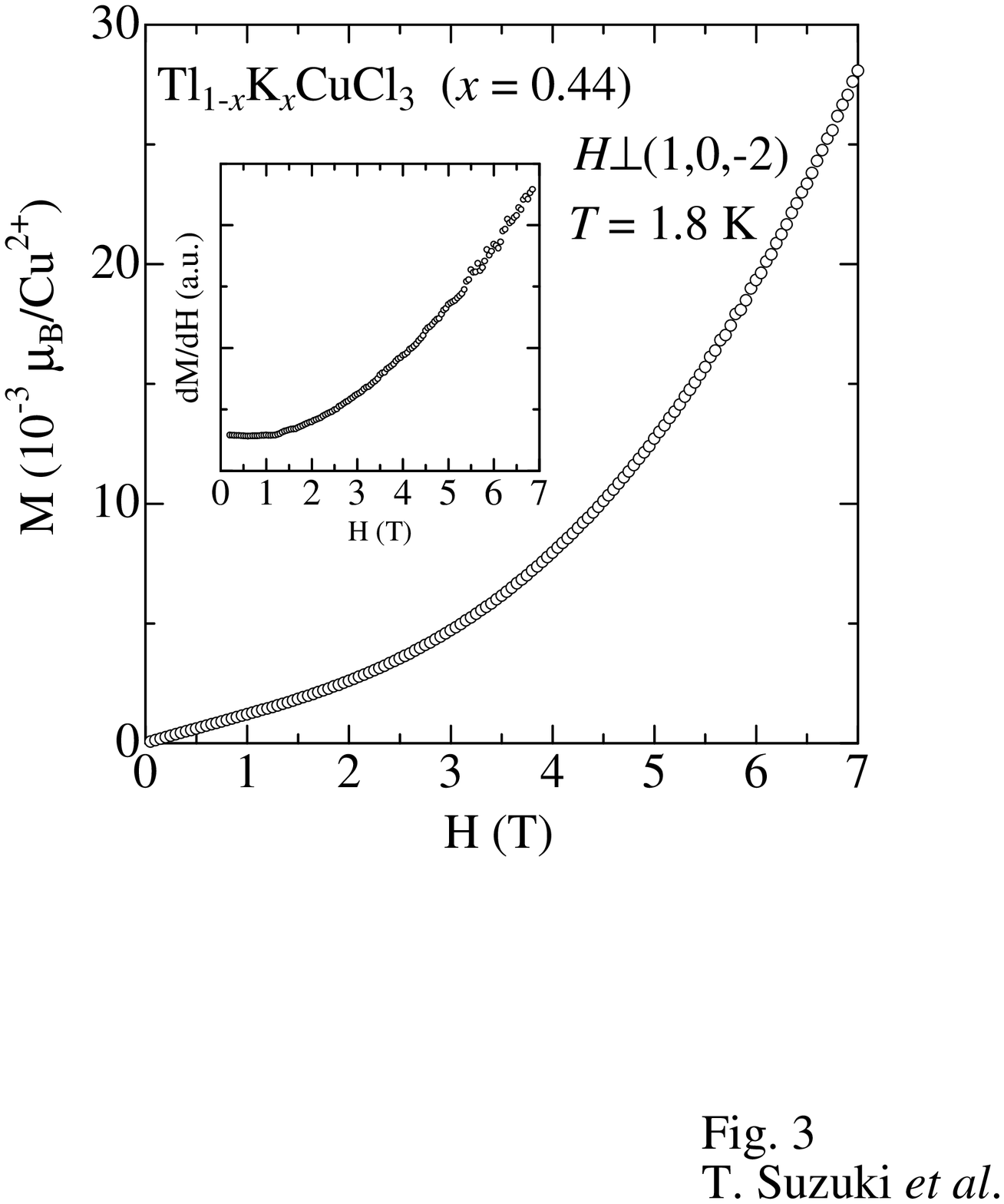}}
\caption{
Magnetic field dependence of the magnetization at 1.8 K.
The inset shows the magnetic field dependence of $dM/dH$.
}
\end{figure}
It is suggested that the strong randomness is introduced intrinsically in the single crystals with the concentration of $x$ = 0.44, although it is not clear whether an excitation gap $\Delta$ develops.
From these magnetization results, the spin system is not in a magnetically ordered state but in a paramagnetic state with a finite susceptibility at the zero-field limit, at least down to 1.8 K.
\\
\indent
Figure 4 shows ZF-$\mu$SR time spectra of Tl$_{1-x}$K$_x$CuCl$_3$ with $x =$ 0.44 at each temperature.
The shape of the time spectrum is changed drastically with decreasing temperature.
The spectrum shape at 10 K is a Gaussian-like function (opens downwards).
At lower temperatures, the spectrum becomes exponential-like at 3 K, and finally, a rather fast relaxation is observed at 80 mK.
\begin{figure}[htb]
\scalebox{0.5}{\includegraphics*[20mm,85mm][200mm,260mm]{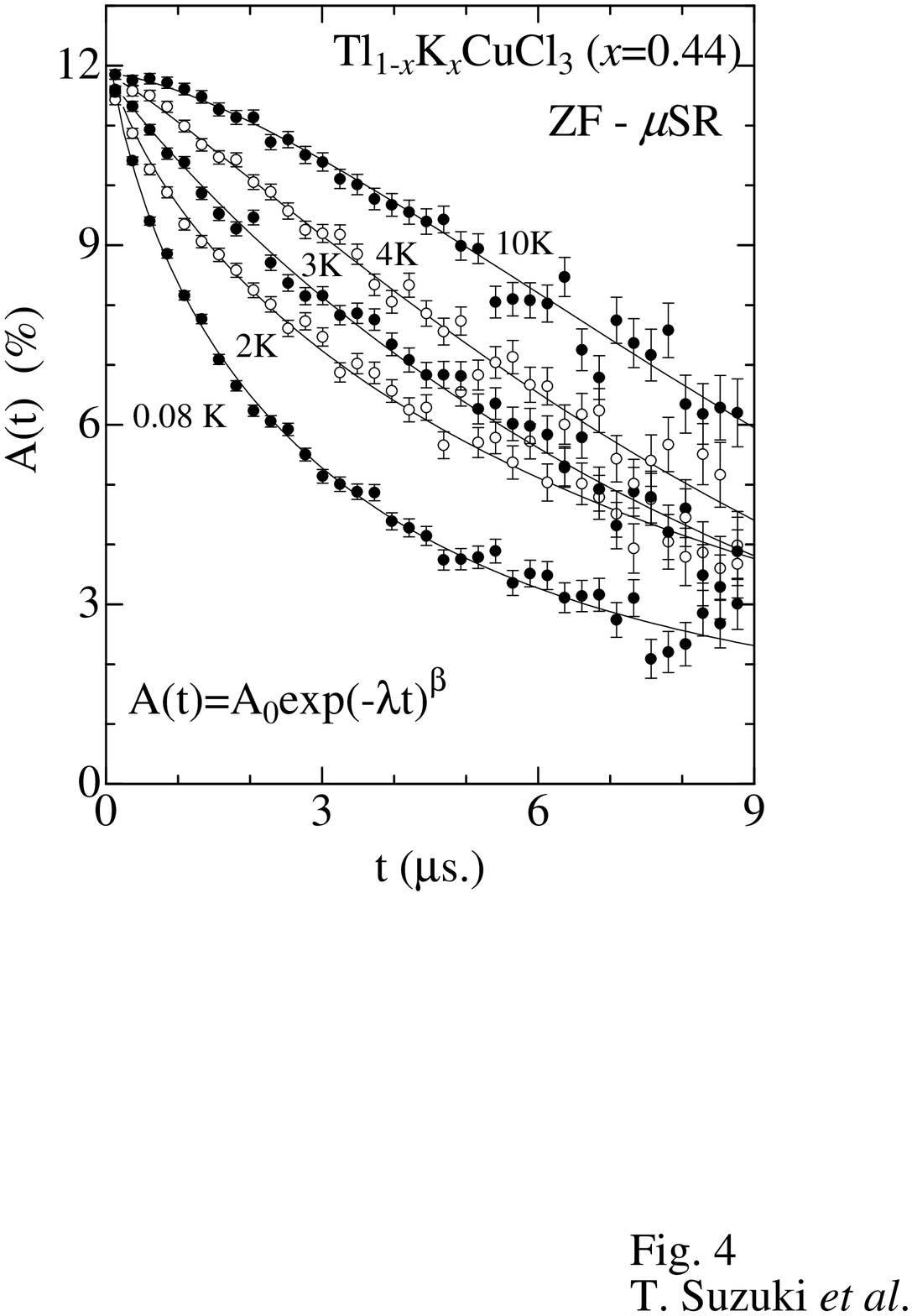}}
\caption{
Time spectrum of the zero-field muon-spin-relaxation (ZF-$\mu$SR) of Tl$_{1-x}$K$_x$CuCl$_3$ with $x$ = 0.44 at each temperature.
Solid lines are fitted results using the stretched exponential function $\exp (-\lambda t)^{\beta}$.
}
\end{figure}
In order to discuss the spectrum change by one formula in the whole temperature range in this study, the $\mu$SR time spectra are analyzed using the function of the stretched exponential $A_{\rm 0}\exp (-\lambda t)^{\beta}$, where $A_{\rm 0}$ is the initial asymmetry and $\lambda$ is the muon-spin-relaxation rate.
All the time spectra are well fitted by the above function, as shown in Fig.4 with solid lines.
\\
\indent
Figure 5 shows the temperature dependence of the muon-spin-relaxation rate $\lambda$ and of the power $\beta$.
The relaxation rate $\lambda$ increases rapidly with decreasing temperature in contrast to the case of $x =$ 0.20 (open triangles).
The rapid increase of the relaxation rate $\lambda$ reminds us of the critical divergence toward a phase transition, and it is suggested that there exists a critical slowing down of the fluctuation frequency of the Cu-3{\it d} spins to a spin frozen state below 80 mK in zero-field.
The expected spin frozen state is not the Bose-glass phase, because the saturation of the relaxation rate $\lambda$ is not observed in the case of $x =$ 0.44, as mentioned below.
\begin{figure}[htb]
\scalebox{0.48}{\includegraphics*[20mm,100mm][220mm,260mm]{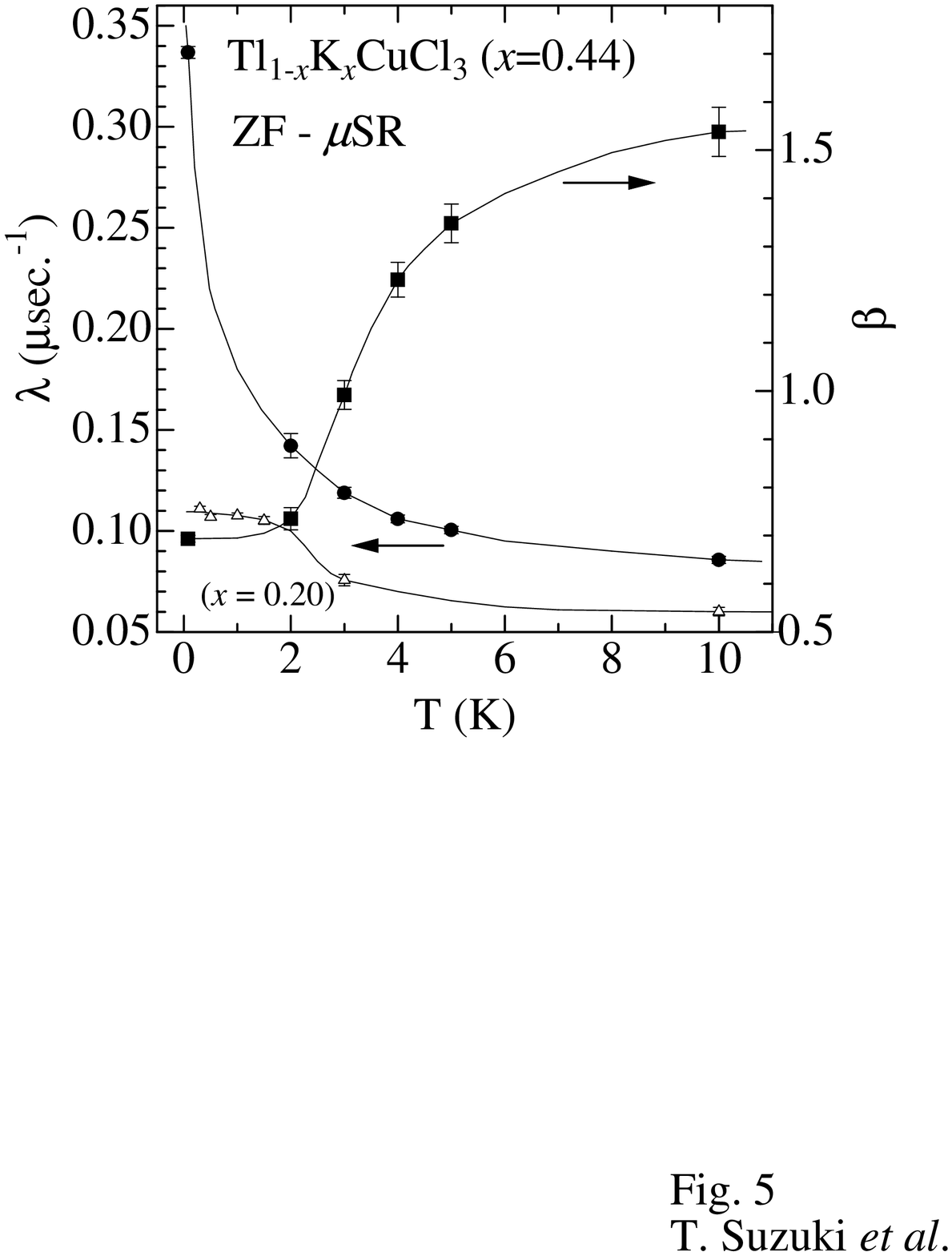}}
\caption{
Temperature dependence of the relaxation rate $\lambda$ (the left side vertical axis, closed circles) and of the power $\beta$ (the right side vertical axis, closed squares) of Tl$_{1-x}$K$_x$CuCl$_3$ with $x$ = 0.44.
Open triangles denote the relaxation rate $\lambda$ for $x$ = 0.20 (ref.38).
Solid lines are guides for the eye.
}
\end{figure}
The power $\beta$ of the stretched exponential function decreases with decreasing temperature, and tends to saturate to the value of $\beta =$ 0.65.
\\
\indent
McHenry {\it et al}. discussed the nuclear spin-lattice relaxation by spatially localized impurity spins in intermetallic compounds~\cite{McHenry}.
According to their results, if magnetic moments which contribute to the spin-lattice relaxation distribute densely, the spin relaxation curve is expressed by the exponential function of $\exp (-\lambda t)$ $(\beta = 1)$ regardless of whether magnetic moments distribute homogeneously in space or not.
For low impurity concentration, however, the spin relaxation curve is calculated to be proportional to $\exp (-\lambda t)^{1/2}$ $(\beta = 1/2)$ when the nuclear-spin diffusion is absent and the localized spins are fluctuating.
\\
\indent
Since the muon-spin-relaxation is caused by fluctuating local magnetic fields at muon sites induced by surrounding magnetic moments, we can apply these results of the nuclear-spin-relaxation analysis to the case in the $\mu$SR measurement.~\cite{Fiory,Uemura2}
Although it is difficult to distinguish whether or not spins are fluctuating without longitudinal-field muon-spin-relaxation measurements, it can be suggested from the root-exponential-like ($\beta =$ 0.65) behavior observed at 80 mK in this study that the fast muon-spin-relaxation originates from the spatially-fixed dilute moments fluctuating in time , and it is suggested that spatially separated islands, which have a finite magnetic moment large enough to cause the fast muon-spin-relaxation, appear in the singlet sea at lower temperatures.
\\
\indent
Figure 6 shows the time spectrum of the ZF-$\mu$SR in Tl$_{1-x}$K$_x$CuCl$_3$ with $x$ = 0.20 at 0.3 K and with $x$ = 0.44 at 80 mK.
\begin{figure}[hhh]
\scalebox{0.5}{\includegraphics*[20mm,82mm][220mm,235mm]{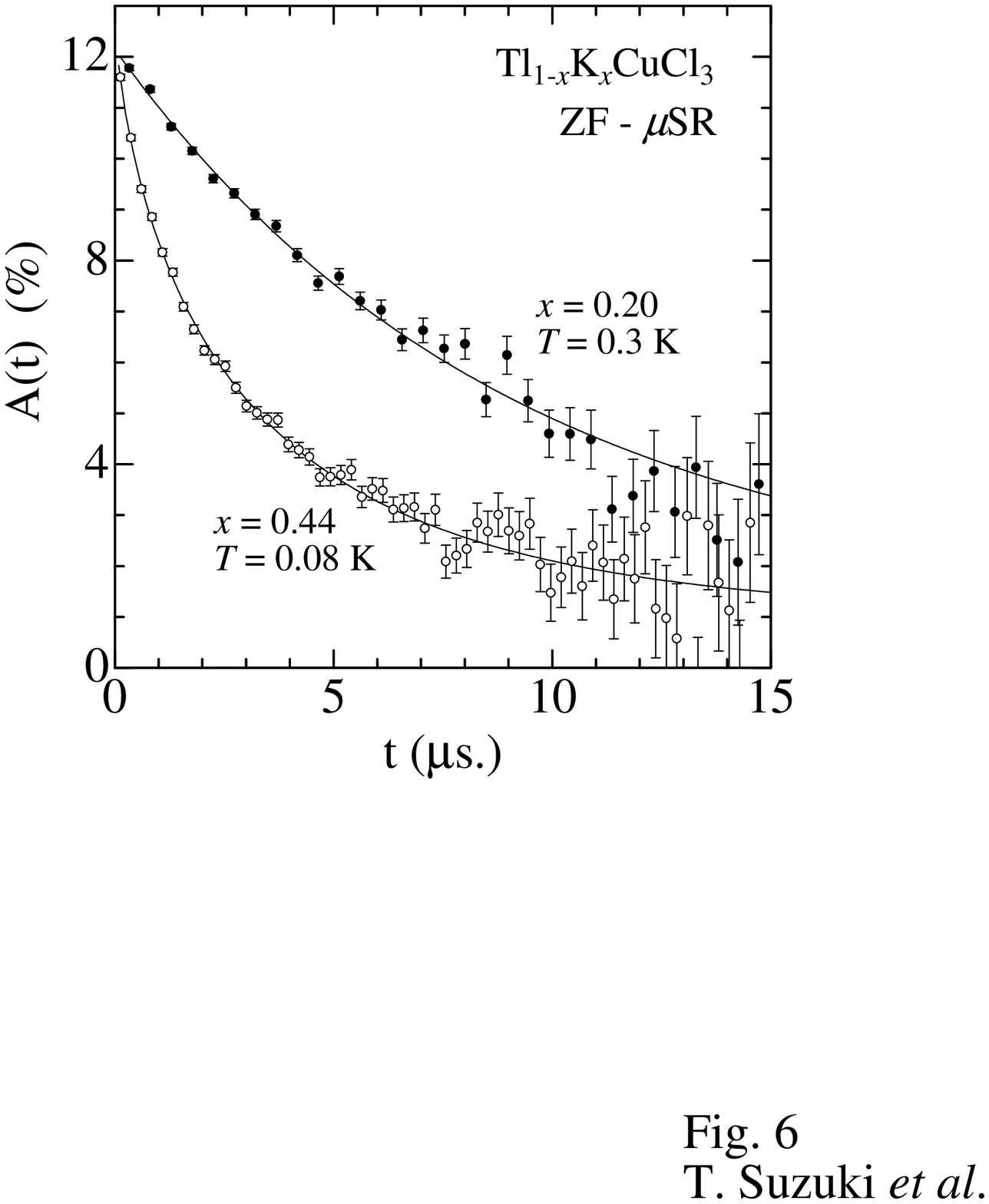}}
\caption{
Time spectrum of the zero-field muon-spin-relaxation (ZF-$\mu$SR) of Tl$_{1-x}$K$_x$CuCl$_3$ with $x$ = 0.20 at 0.3 K (ref.38) and $x$ = 0.44 at 80 mK.
}
\end{figure}
In the case of $x =$ 0.20, the time spectrum is well fitted by a simple exponential function, and the muon-spin-relaxation rate $\lambda$ tends to saturate at lower temperatures as shown in Fig.5~\cite{Suzuki}.
The saturation of $\lambda$ in the case of $x =$ 0.20 is consistent with the existence of the Bose-glass phase at {\it T} = 0 by analogy with the result on the frustrated system which has quantum spin fluctuations without static local magnetic fields at muon sites down to 100 mK~\cite{Uemura}.
In the case of $x =$ 0.44, however, the shape of the time spectrum is the root-exponential-like, and the relaxation rate $\lambda$ shows the rapid increase with decreasing temperature as mentioned above.
This result observed in $x =$ 0.44 is quite different from the case in $x =$ 0.2, and this difference indicates the ground state changes with increasing the concentration $x$ corresponding to the degree of the bond-randomness.
The existence of a new ordered phase is expected in highly random regions.
Finally, we emphasize that the result reported in this study is a new phenomenon which has not been predicted theoretically or observed previously in this system.

\section{Summary}

\indent
In summary, we carried out magnetization and zero-field muon-spin-relaxation (ZF-$\mu$SR) measurements in the bond-randomness introduced quantum spin system Tl$_{1-x}$K$_x$CuCl$_3$ with $x = $ 0.44 single crystals.
The muon-spin-relaxation rate $\lambda$ increases rapidly with decreasing temperature, and the time spectrum at 80 mK shows a root-exponential like behavior.
These results suggest that below 80 mK, in contrast to the predicted Bose-glass phase, there exists a critical slowing down of the frequency of the Cu-3{\it d} spin fluctuation toward a spin frozen state.
The muon-spin-relaxation $\lambda$ in this state originates from spatially-fixed fluctuating dilute moments.

\section*{Acknowledgment}
The authors are grateful to Dr. F. Pratt for a critical reading of the manuscript.
This research was partially supported by the Joint Research Projects of JSPS, and partially supported by the Torey Science Foundation.

\end{document}